\documentclass[twocolumn,showpacs,preprintnumbers,prl]{revtex4}
\usepackage{graphicx}
\usepackage{dcolumn}
\usepackage{bm}
\usepackage{amsmath}
\newcommand{\kagome}{kagom\'{e} }
\newcommand{\czodcx}{$\rm Cu_{4-x}Zn_x(OD)_6Cl_2$ }
\newcommand{\czodc}{$\rm Cu_{3}Zn(OD)_6Cl_2$ }
\newcommand{\codc}{$\rm Cu_{2}(OD)_3Cl$}
\newcommand{\neel}{N\'{e}el }
\newcommand{\mbulk}{$\langle M\rangle_{bulk}$}
\newcommand{\dmbulk}{$d\langle M\rangle_{bulk}/dH$}

\begin{document}
\preprint{}
\title{External magnetic field effects on a distorted kagome antiferromagnet}
\author{J.-H. Kim$^{1}$}
\author{S. Ji$^{1,2}$}
\author{S.-H. Lee$^{1}$}
\author{B. Lake$^{3,4}$}
\author{T. Yildirim$^2$}
\author{H. Nojiri$^5$}
\author{H. Kikuchi$^6$}
\author{K. Habicht$^3$}
\author{Y. Qiu$^2$}
\author{K. Kiefer$^3$}
\affiliation{$^1$Department of Physics, University of Virginia, Charlottesville, Virginia 22904\\
$^2$NIST Center for Neutron Research, National Institute of Standards and Technology, Gaithersburg, Maryland 20899\\
$^3$Hahn-Meitner-Institut, Glienicker Strabe 100, Berlin D-14109, Germany\\
$^4$Institut f?r Festk?rperphysik, Technische Universit?t Berlin, Hardenbergstr. 36, 10623 Berlin, Germany\\
$^{5}$Institute for Materials Research, Tohoku University, Sendai, Miyagi 980-0821, Japan\\
$^6$Department of Applied Physics, University of Fukui, Fukui 910-8507, Japan
}

\date{\today}

\begin{abstract}
We report bulk magnetization, and elastic  and inelastic neutron
scattering measurements under an external magnetic field, $H$, on
the weakly coupled distorted \kagome system, \codc. Our results
show that the ordered state below 6.7 K is a canted
antiferromagnet and consists of large antiferromagnetic
$ac$-components and smaller ferromagnetic $b$-components.  By
first-principle calculations and linear spin wave analysis, we
present a simple spin hamiltonian with non-uniform nearest
neighbor exchange interactions resulting in a system of coupled
spin trimers with a single-ion anisotropy that can
qualitatively reproduce the spin dynamics of \codc.

\end{abstract}


\maketitle

The quest for novel quantum spin states in a space higher than
one-dimension has been active \cite{Sachdev08} since the
resonating valence bond state was proposed as the ground state of
a two-dimensional quantum spin triangular system \cite{Anderson73}
and was also proposed to be relevant to high $T_c$
superconductivity in cuprates \cite{Anderson87}. Recently, a few
compounds have attracted lots of attention as good model systems
\cite{Levi07}. One of them is \czodcx which was initially believed
to realize a two-dimensional kagome lattice with quantum spins
when x = 1 and a three-dimensional pyrochlore when x =
0.\cite{Shores05} The prospect looked promising because \czodcx~
does not order down to 50 mK for most values of $x$.
\cite{Helton07}

Recent neutron diffraction studies, however, have found that
\czodc with rhombohedral $R\bar{3}m$ symmetry realizes a perfect
kagome lattice but with ~ 10\% nonmagnetic site
disorder.\cite{shl07, vries08} Furthermore, when Zn ions are
replaced by Cu ions, \czodcx with $x < 0.3$, becomes monoclinic
($P2_1/n$) due to the Jahn-Teller effects of Cu$^{2+} (3d^9)$ ion.
This leads to different bond angles for different Cu-O-Cu
superexchange paths in such a way that the kagome lattice is
distorted to have nonuniform nearest neighbor interactions and the
interplane coupling between the kagome and triangular planes is
expected to be weak. Thus \czodcx realizes a good two dimensional
quantum spin system for all $x$ but it does have defects with
regards to an ideal kagome quantum antiferromagnet: for $x = 1$
the nonmagnetic site disorder while for $x = 0$ the nonuniform
coupling constants.\cite{shl07}

The distorted kagome quantum spin system \codc\ is interesting in
its own right. Specific heat measurements have shown that upon
cooling it undergoes two transitions, one at $T_{c2} =$ 18 K and
another at $T_{c1} =$ 6.7 K. \cite{Zheng05} Muon spin relaxation
($\mu$SR) measurements yielded evidence for long range order at
$T_{c2}$ and for enhancement of strong spin fluctuations at
$T_{c1}$. \cite{Zheng05, Mendels07} However, in the specific heat
data the amount of entropy given off at the $T_{c2}$ transition is
tiny in comparison to the change in entropy at $T_{c1}$, which
appears inconsistent with the $\mu$SR results. Neutron diffraction
studies revealed development of observable magnetic Bragg peaks
only below $T_{c1}$ \cite{shl07}, which indicates that this
transition is associated with \neel~order, a result that is
consistent with the large amount of entropy released at $T_{c1}$
but not at $T_{c2}$. The spin structure was found to be collinear
rather than the 120$^o$ configuration that is expected for a
uniform kagome antiferromagnet. A recent theoretical study showed
that collinear spin structures can indeed be favored by a
distorted kagome.\cite{Lawler08} For $T_{c1} < T < T_{c2}$, on the
other hand, there were no observable magnetic peaks but instead a
weak magnetic excitation was observed at an energy transfer
$\hbar\omega =$ 7 meV.\cite{shl07} This intermediate phase was
proposed to be a quantum spin singlet state, in contradiction with
the $\mu$SR results.

Here we present our bulk magnetization and neutron scattering
measurements on \codc~under an external magnetic field. Our
principal results are the following.
(1) The long-range magnetic order below 6.7K consists of strong
antiferromagnetic collinear $ac$-components ($\mid \langle
M\rangle_{ac}\mid = 0.56(10)$ $\mu_B$/Cu$^{2+}$) and a weak
ferromagnetic $b$-component ($\mid \langle M\rangle_{b}\mid =
0.151$ $\mu_B$/Cu$^{2+}$). (2) The 7 meV peak in the magnetic
excitation spectrum, that was previously ascribed to a quantum
singlet state, is due to a van Hove singularity of the spin wave
excitations at the zone boundary of the excitation spectrum. (3)
The spin dynamics can be qualitatively explained by a simple spin
hamiltonian with non-uniform nearest neighbor exchange
interactions resulting in three different coupling constants with
values of 8 meV, 3.7 meV and 2.8 meV in the \kagome plane and a single ion
anisotropy of 0.4 meV.

A 20 g powder sample of \codc\ was made at the University of Fukui
using the standard method reported elsewhere.\cite{Shores05}  Bulk
magnetization measurements up to H = 12 T were performed at the
Hahn-Meitner Institut (HMI), Berlin, while high field measurements
up to H = 35 T were done at the IMR, Tohoku University. Neutron
scattering measurements were performed at the NIST Center for
Neutron Research (NCNR) and at the Berlin Neutron Scattering
Center (BENSC) at HMI. Details of the neutron scattering
experimental setups are described in the captions of the figures
that present the data.

\begin{figure}
\includegraphics[width=\hsize]{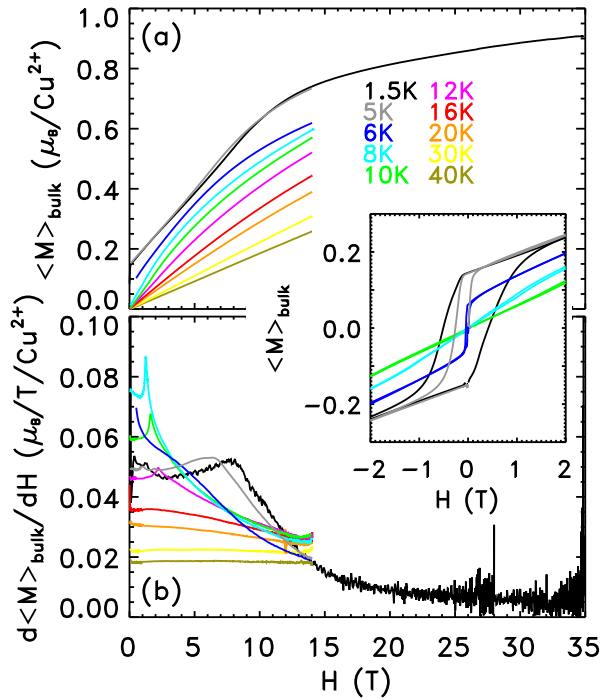}
\caption{(a) Bulk magnetization, \mbulk, and (b) susceptibility,
\dmbulk, measured as a function of an external magnetic field at
several different temperatures. The inset shows the low field (-2
T $< H <$ 2 T) data of \mbulk~at five selected temperatures.}
\label{refine}
\end{figure}
\begin{figure}
\includegraphics[width=\hsize]{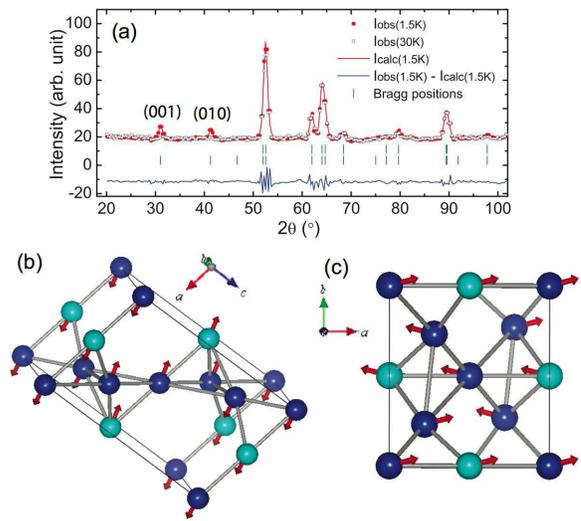}
\caption{(a) Elastic neutron scattering as a function of the
scattering angle, $2\theta$. The spin structure projected (b) on
the $ac$-plane and (c) on the $ab$-plane. Blue and green spheres in (b) and (c) represent \kagome~and triangular Cu$^{2+}$ spins, respectively.} \label{refine}
\end{figure}

Fig. 1 (a) shows bulk magnetization, \mbulk, measured as a
function of an external magnetic field, $H$, at various
temperatures, spanning the two phase transitions at $T_{c1} =$ 6.7
K and $T_{c2} \sim$ 20 K. At $T = 1.5$ K $<< T_{c1}$, \mbulk~has a
nonzero value of 0.15 $\mu_B/$Cu$^{2+}$ at $H =$ 0 Tesla and
exhibits hysteresis with $H$ (see inset). This tells us that the
\neel~state below $T_{c1}$ has ferromagnetic components. As shown
in Fig. 2 (a), below $T_{c1}$, antiferromagnetic Bragg reflections
with the characteristic wave vector of ${\bf Q}_m$ = (0,0,0)
appear. The ferromagnetic component clearly observed in
\mbulk~would yield magnetic scattering on the top of the nuclear
Bragg reflections only. We have applied the group theoretical
approach in solving the magnetic structure. There are four
irreducible representations of group $P2_1/n$ for ${\bf Q}_m$.
Among them, the representation $\tau_1$ has  antiferromagnetic
basis functions for the crystallographic $a$- and $c$-components,
$\psi_{\tau_1}^a = (1,0,0)$ and $\psi_{\tau_1}^c = (0,0,1)$,
respectively, and ferromagnetic $b$-component, $\psi_{\tau_1}^b =
(0,1,0)$. We have found that the linear combination of the basis
functions of the $\tau_1$ representation, $0.55(5) \psi_{\tau_1}^a
+ 0.151\psi_{\tau_1}^b+0.23(25) \psi_{\tau_1}^c$, reproduces our
elastic neutron scattering pattern (Fig. 2 (a)). As shown in Fig.
2 (b), the $ac$-components form the collinear, antiferromagnetic
spin arrangement that was proposed previously. \cite{shl07}

When an external magnetic field is applied up to about 10 Tesla,
\mbulk~increases rapidly to the average value of 0.7
$\mu_B/$Cu$^{2+}$. For H $>$ 10 Tesla, \mbulk~ keeps increasing at
a slower rate to 0.9 $\mu_B/$Cu$^{2+}$ at $H=$ 35 Tesla which is
close to the value of 1 $\mu_B$/Cu$^{2+}$, expected when all spins
are fully polarized ferromagnetically. The change in the increase
rate can also be seen in \dmbulk~vs $H$ (Fig. 1 (b)) which
exhibits a broad peak at around $H =$ 8 Tesla and decreases
rapidly at higher $H$, indicating a field-induced magnetic phase
transition. Upon warming, the antiferromagnetic $ac$-components
weaken\cite{shl07}, and so does the ferromagnetic $b$-component
(see the 6 K data in the inset), and both components disappear at
$T_{c1} =$ 6.7 K\cite{Shores05,shl07}. At $T =$ 6 K $\sim T_{c1}$,
the broad peak which was found in \dmbulk~at $H_{c1} \sim$ 8 Tesla
at $T << T_{c1}$ becomes a broader bump at a lower $H_{c1} \sim 5$
Tesla. While in the intermediate phase, $T_{c1} < T < T_{c2}$, the
broad peak found in \dmbulk~for $T < T_{c1}$ is now replaced by a
sharp peak at low $H$; $H_{c2} =$ 1.5, 2, 2.2 Tesla for $T=$ 8,
10, 12 K, respectively, and at 16 K a broad peak at around $H_{c2}
=$ 2 Tesla. At 30 K and 40 K $ > T_{c2}$, \dmbulk~is nearly flat,
i.e., \mbulk~increases linearly with $H$, up to $H = 14$ Tesla,
suggesting paramagnetic behaviour.

\begin{figure}
\includegraphics[width=7cm]{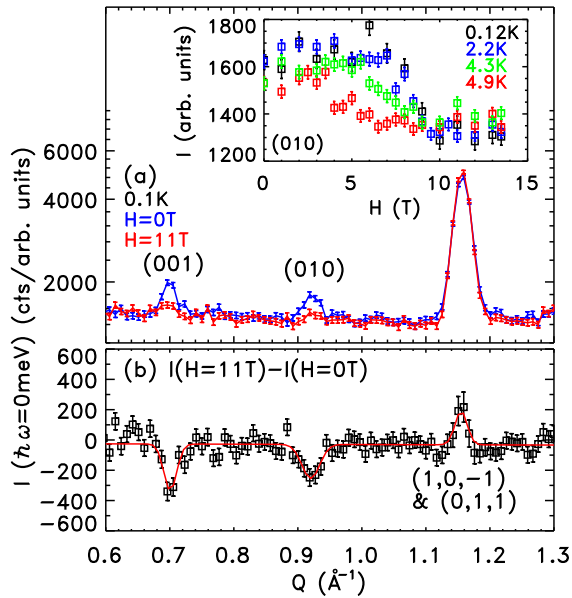}
\caption{(a) Elastic neutron scattering pattern as a function of
$Q$ up to 1.3 $\AA^{-1}$, obtained at $T = 0.1$ K under an
external magnetic field of  0 T (blue) and 3.5 T (red). (b) The
difference between the two diffraction pattern, I (H = 11 T) $-$ I
(H = 0 T). The data were obtained with an incident neutron energy
of $E_i = 3.7$ meV and collimation of guide-80-80-open at the
cold-neutron triple-axis spectrometer, SPINS, located at the NCNR.
Contamination from higher-order neutrons were eliminated by
placing a BeO filter after the sample. Inset of (a): H-dependence
of the (010) Bragg peak intensity at four different temperatures,
T = 0.12 K, 2.2 K, 4.3 K and 4.9 K. The measurements were done at
the cold neutron triple-axis spectrometer V2 at HMI with $E_i =$ 5
meV.} \label{refine}
\end{figure}

To study the field effects on the Neel state below $T_{c1}$, we
have performed elastic neutron scattering measurements with an
external magnetic field. Fig. 3 (a) shows the elastic neutron
scattering pattern as a function of momentum transfer, $Q$,
measured at $T$ = 0.12 K with two different fields, $H =$ 0 T
(blue symbols) and 11 T (red symbols). The two antiferromagnetic
Bragg peaks at the (001) and (010) reflections, that are prominent
in zero field, become very weak at $H$ = 11 T, indicating a field
induced phase transition occurred below $H=$ 11 T. The inset of
Fig.\ 3 (a) shows that at $T=$ 0.12 K the peak starts weakening at
around 7 T which is close to the $H_{c1} = 8$ T observed in
\dmbulk~ and disappears at around 10 T. When temperature
increases, the critical field decreases, for instance $H_{c1} =$
6.5 T at $T=$ 4.9 K, which is consistent with our bulk
magnetization data. Where did the antiferromagnetic scattering go?
The intensity difference between $H=0$ and 11 T, plotted in Fig. 3
(b), shows that the peak at $Q = 1.16 \AA^{-1}$ which corresponds
to the nuclear (1,0,-1) $\&$ (0,1,1) reflections has increased
upon application of the field. Altogether, the neutron and
susceptibility data suggest a picture where below 8T the field
increases the size of the spin moment without changing its
direction significantly. In contrast, above 8T where the
ferromagnetic component increases at the expense of the
antiferromagnetic component and the spins are able to increase
their alignment with the field perhaps by overcoming an
anisotropy.

\begin{figure}
\includegraphics[width=6cm]{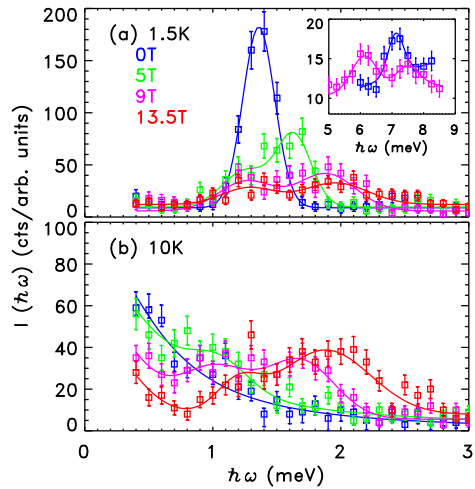}
\caption{Inelastic neutron scattering intensity obtained at $Q =
0.6$ \AA$^{-1}$ under external magnetic fields of, H = 0 T (blue),
5 T (green), 9 T (pink) and 13.5 T (red), at temperatures of T =
1.5 K (a) and 10 K (b). The data were obtained at the cold-neutron
triple-axis spectrometer V2.
The data shown in the inset of (a) were obtained at $Q = 1.2$
\AA$^{-1}$, at the SPINS. At both spectrometers, the energy
of the scattered neutrons was fixed to be $E_f = 3.7$ meV and a
horizontally focusing analyzer was utilized to increase the
sensitivity.} \label{refine}
\end{figure}

Let us now turn to field effects on spin fluctuations. Fig. 4 (a)
shows the H-dependence of the magnetic excitation spectrum
measured at $T=$ 1.5 K $<<T_{c1}$. At $H = 0$ Tesla, there are two
modes centered at $\hbar\omega =$ 1.32 meV and 7 meV, as
previously reported \cite{shl07}. When H is applied, the sharp
peak at 1.3 meV splits into two broad peaks and the splitting
between the two peaks increases with H. At H = 13.5 T, the two
peaks are centered at $\hbar\omega=$ 1.25 meV and 2 meV,
respectively.
The 7 meV peak also splits into two peaks  (see inset), which
unambiguously tells us that it is not due to the
singlet-to-triplet excitations of a quantum dimer as previously
thought, but is due to spin waves. At 10 K $> T_{c1}$, and zero
field, the $\hbar\omega=$ 1.3 meV mode is replaced by a low energy
continuum (see Fig. 4 (b)), commonly found in a cooperative
paramagnet. When the magnetic field is applied, distinct
excitation modes appear out of the continuum. For $H \geq$ 9 T,
there exist a broad feature that can be fitted by two gaussians.
At $H$ = 13.5 T, the two peaks are centered at ~ 1.25 meV and 2
meV. This behavior is quite different from the field-induced
single-ionic Zeeman-like behavior observed in \czodcx (x=0.66 and
1)\cite{Helton07, shl07}. Instead, it is very similar to the
behavior observed at 1.5 K (see Fig. 4 (a)) in an external field.
This indicates that the paramagnetic continuum at low energies
above $T_{c1}$ is due to spin fluctuations anticipating the canted
antiferromagnetic order below $T_{c1}$. This also suggests that
the 7 meV mode that survives above $T_{c1}$ might be related to
the spin wave fluctuations of the \neel~state.

\begin{figure}
\includegraphics[width=\hsize]{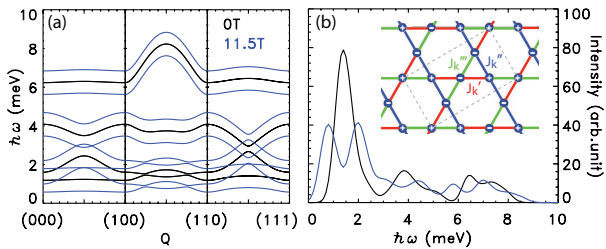}
\caption{(a) Dispersion of spin wave excitations along (h,0,0),
(1,k,0), and (1,1,l) directions. (b) Powder-averaged neutron
scattering intensity as a function of $\hbar\omega$. Black and
blue lines are the results for H = 0T and 11.5 T, respectively.
The inset of (b) shows the \kagome lattice with the ac-components
of the moments. Red, blue, green lines represent different bond
strengths.} \label{refine}
\end{figure}

It is almost impossible to extract the exact spin hamiltonian,
${\cal H}$, from powder-averaged inelastic neutron scattering
data, because the crystallographic directional information is lost. 
This is especially the case when ${\cal H}$ is complex as in
\codc. Therefore we have performed the local-density approximation (LDA)+$U$ (the $d-d$ Coulomb interaction parameter) calculations for a wide range of $U$ to
obtain the ratios of different exchange constants. Our results
tell us that there are three different coupling constants in the
\kagome plane as shown in the inset of Fig. 5b with the ratios of
$J_{k}^{''}/J_{k}^{'} \simeq 0.46$ and $J_{k}^{'''}/J_{k}^{'} \simeq 0.35$.\cite{taner} 
The inter-plane couplings turned out to be much weaker, less than $0.1 J_k^{'}$. We
then performed linear spin wave analysis  with those ratios fixed
and with two fitting parameters, $J_{k}^{'}$, that couples the
spins into trimers, and a single ion anisotropy along the spin
direction, $D$, for the collinear AFM structure composed of the $ac$-components of the Cu$^{2+}$ ions, neglecting the FM $b$-components.

We focused on the following experimental observations: upon
warming the 1.3 meV mode shifts to lower energies as the frozen
moment weakens while the 7 meV mode does not shift in
energy.\cite{shl07} This suggests that the 1.3 meV excitation is a
zero energy mode shifted to finite energy by anisotropy, while the
7 meV mode is due to the van Hove singularity of the excitations
at the zone boundary of the excitation spectrum. 
As shown in Fig. 5 (b), a sharp and strong peak at 1.3 meV and a
weak and broad peak at 7 meV can be generated in zero field with
$J_{k}^{'} = 8$ meV and $D = -0.4$ meV.
Fig. 5 (a) show that the 1.3 meV peak is due to two nearly-flat doubly
generate modes, while the 7 meV peak
arises from the enhanced density of states at the zone-boundary
energy of another doubly degenerate mode. The fourth doubly degenerate mode generates two broad
peaks around 4 meV, which have not been observed experimentally.
The discrepancy may tell us that the exchange interactions are
not isotropic or the asymmetric exchange interactions such as the Dzyaloshinskii-Moriya interactions need to be included in our analysis.

When a Zeeman term, $g\mu_B H\cdot \sum S_{i}^a$, is included in
the spin hamiltonian, the 1.3 meV peak splits into two peaks at a
lower and a higher energy, while the excitations around 7 meV
becomes broader and exhibit two peaks at around 6 meV and 7 meV,
which are qualitatively consistent with the data shown in Fig. 4
(a). Thus, the previous interpretation of the 7 meV excitation as
a triplet associated with spin dimers in the
intermediate temperature (6.7 K $<$ T $<$ 18 K) regime
\cite{shl07} should be modified to reflect the large intratrimer interactions. So far, the observed
characteristics of this phase are the weak anomaly in the specific
heat, $\mu$SR measurements, and the small kinks in our \dmbulk~vs
H data. Further experimental and theoretical investigations are
necessary to understand the nature of the intermediate T phase.

In summary, using bulk susceptibility, and elastic and inelastic
neutron scattering measurements with and without an external
magnetic field, we have characterized the static and dynamic spin
correlations in \codc. Our analysis shows that \codc~should be
considered as a system of coupled spin trimers on the \kagome
lattice.

\begin{acknowledgments}

We thank A. B. Harris, S. Nagler, T. J. Sato, Y. B. Kim, M. Kofu for helpful discussion.
\end{acknowledgments}

\end{document}